# The interplay between phase-separation and gene-enhancer communication: a theoretical study

Andrea M. Chiariello[1,2,*], Federico Corberi[1], Mario Salerno[1]

[1]*Dipartimento di Fisica "E.R Caianiello" and INFN, Gruppo Collegato di Salerno, Università di Salerno, via Giovanni Paolo II 132, 84084 Fisciano (SA), Italy.*
[2]*Dipartimento di Fisica "Ettore Pancini", Università degli Studi di Napoli Federico II, and INFN Sezione di Napoli, Complesso Universitario di Monte Sant'Angelo, 80126, Naples, Italy*

[*]*Corresponding author:* <u>chiariello@na.infn.it</u>

**Abstract**

The phase-separation occurring in a system of mutually interacting proteins that can bind on specific sites of a chromatin fiber is here investigated. This is achieved by means of extensive Molecular Dynamics simulations of a simple polymer model which includes regulatory proteins as interacting spherical particles. Our interest is particularly focused on the role played by phase-separation in the formation of molecule aggregates that can join distant regulatory elements, such as gene promoters and enhancers, along the DNA. We find that the overall equilibrium state of the system resulting from the mutual interplay between binding molecules and chromatin can lead, under suitable conditions that depend on molecules concentration, molecule-molecule and molecule-DNA interactions, to the formation of phase-separated molecular clusters allowing robust contacts between regulatory sites. Vice-versa, the presence of regulatory sites can promote the phase-separation process. Different dynamical regimes can generate the enhancer-promoter contact, either by cluster nucleation at binding sites or by bulk spontaneous formation of the mediating cluster to which binding sites are successively attracted. The possibility that such processes can explain experimental live-cell imaging data measuring distances between regulatory sites during time is also discussed.

**Significance**

Phase-separation is a general physical mechanism that occurs in living cells at various levels and is fundamental for genome activity. Indeed, protein condensates mediate the interaction between distant regulatory elements along chromatin chain and shape genome architecture. Correspondingly, the activity of the genes is strongly associated with this process. Remarkable experimental work has been recently done to investigate phase-separation and it is currently object of intense research. Here, using polymer modelling and Molecular Dynamics simulations we provide a systematic investigation of this fundamental process and explore its influence on the physical communication between regulatory elements along the chromatin chain, as genes and enhancers.

**INTRODUCTION**

The formation of molecular aggregates through phase-separation (PS) is a general physical mechanism for which there is an increasing amount of experimental evidence highlighting its importance for the cell activity (1, 2). Within the cell nucleus, PS occurs at many levels and leads to the formation of membraneless structures at different scales, ranging from micron-sized aggregates as nucleoli and Cajal bodies (1, 2) to the few hundreds of nano-meters droplets of transcriptional coactivators (3), mediator and RNA PolII clusters (4, 5) and transcription factors(6). Typically, phase-separated aggregates have round shapes, tend to coalesce if in spatial proximity and exhibit a highly dynamic, liquid-like behaviour, as highlighted by photobleaching experiments (3, 5). Importantly, the formation of such condensates is deeply linked to the transcriptional activity of the genes, as they mediate the contact with their distal regulatory elements, that is enhancers and super-enhancers (7). On the other hand, PS plays also a crucial role in shaping repressed heterochromatin through the formation of phase-separated condensates of the protein HP1 (8, 9). Recently, it has been shown that the chromatin fiber can undergo PS forming dense chromatin droplets, as shown in *in-vitro* experiments at physiological conditions(10). Also, the phase-separated aggregates have been shown to have a complex, multi-layered internal structure emerging from the different interactions among the molecules (11).

In contrast with prokaryotic cells, where the organization of the DNA is much simpler and the mechanism of gene activation may involve the sliding of regulatory proteins along the DNA (12), in eukaryotic cells the situation is much more complex and the PS process plays an important role either for the spatial organization of chromatin within the nucleus and for gene regulation (13, 14). Indeed, genome structure is intimately linked to the transcriptional activity of genes, as a correct folding allows an efficient communication between genes and their distal enhancers (15), while, if altered, can cause severe diseases (16). To quantitatively investigate such three-dimensional (3D) architecture classic Polymer Physics models (17, 18) and mesoscale models (19) have been developed. Notably, they successfully explained general aspects of genomic structure and helped to better understand mechanistic principles that regulate chromatin folding, as the formation of chromatin loops (20, 21), the structure of mega-base sized human and murine loci (22–24), the impact of structural variants (25) and the structure of real loci at nucleosome level (26). Importantly, some of those models (27–29) rely on simple thermodynamic mechanisms and naturally envisage the formation of phase-separated molecular clusters necessary to mediate the contact between distant elements along the chromatin chain (30).

In this paper, we use Polymer Physics and Molecular Dynamics simulations (28) to quantitatively investigate the relationship between the PS process and the formation of contacts between genes and enhancers. To this aim, we consider a simple model in which specific regulatory loci of the chromatin fiber can interact with diffusing multivalent molecules that in turn non-specifically interact among themselves. Such interactions are known to exist for several proteins and are fundamental to promote PS events (2, 31). From this point of view, the model is a generalization of previous models (32) in which the above interactions were typically overlooked.

By varying the parameters controlling the phase-transition, which are the molecular concentration and the non-specific affinity, we build the system's phase-diagram, extending the usual experimental approach typically focused only on the molecular concentration. The PS process is triggered by increasing the molecular concentration or their interaction affinity above threshold values. When the transition occurs and equilibrium is achieved, the structural properties of the molecular cluster can be changed by means of the control parameters. For very weak affinities, we find that the cluster exhibits dynamical properties such as high exchange rates of particles and internal mobility, indicating a liquid-like nature of the molecular aggregate. Consistently with experiments (3, 6), the formation of the phase-separated cluster is crucial to mediate a stable contact between the enhancer and its target gene. The predictions of the model are then compared with published experimental live-imaging data measuring distances between the *Sox2* promoter and its super-enhancer SCR (33). On the other hand, genes and enhancers can act as nucleation starting site and induce the PS. In general, the formation of molecular aggregates driven by PS and the interaction of the molecules with the genes (or enhancers) located along the chromatin chain act cooperatively and their interplay determine transient and equilibrium properties of the system.

## RESULTS

**The simulated system**

To investigate the PS process mediating the contact between distant loci on the chromatin filament, we consider a model consisting of a polymer where two distant binding regions are located. Such regions can attractively interact with particles (named "binders" in the following) floating in the surrounding environment. Biologically, the polymer represents a chromatin filament and the binding sites represent the enhancer and the promoter of its target gene (Figure 1), while the binders mimic the molecular factors that normally tie to chromatin in the cell nucleus, e.g. Transcription Factors (TFs). Importantly, binders can also attractively interact among each other in a non-specific manner, since it is known that TFs undergo weak attractive interactions through their *intrinsically disordered regions* (IDRs) (1, 2), as highlighted by experimental evidence *in-vitro* (3, 6) and *in-vivo* (5). Thus, the system is regulated by three main parameters: the binders concentration *c*; the interaction affinity between the binders and the DNA binding sites $E_{b-bs}$ and the interaction affinity among the binders $E_{b-b}$ (see Methods). In general, for a fixed value $E_{b-bs}$, if *c* and $E_{b-b}$ are above a certain threshold, a phase-separated cluster of binders can form and mediate a stable contact between distant binding sites (Figure 1A). Details about other specific parameters in the simulations, as temperature and arrangement of the binding regions along the polymer, can be found in the Methods.

**PS process occurs with a switch-like behaviour**

We first study the simple case with a weakly interacting polymer (i.e. with low $E_{b-bs}$ values). Analogous results are obtained for the system in the absence of polymer. To quantitatively study the PS process, we fix the binders concentration and vary the affinity $E_{b-b}$. To evaluate the thermodynamic state of the system, we consider the quantity $N_b/N_{tot}$, where $N_{tot}$ is the total number of binders and $N_b$ is the number of binders contained in the largest of the clusters that spontaneously may form (see Methods). The binding affinity with the polymer is set to $E_{b-bs}$ =3.1$K_B$T, where T is the temperature and $K_B$ the Boltzmann constant (see Methods). As shown in Figure 1B, if the interaction affinity $E_{b-b}$ is below a transition value no macroscopic cluster of binders is observed at equilibrium. Conversely, if the affinity is higher than an energetic threshold, a macroscopic phase-separated cluster is stable at equilibrium. The transition value is identified between $E_{b-b}$ = 2.7$K_B$T and $E_{b-b}$ = 3.0$K_B$T for the considered concentration *c* ≈ 1.1%, where *c* is expressed as volume fraction (see Methods).

Next, we fix the affinity at an above threshold value $E_{b-b}$ = 3.0$K_B$T and vary the binder concentration *c* (Figure 1C, upper panel). Again, the system undergoes a transition and a phase-separated cluster is observed if *c* is above a threshold identified around *c* ≈ 0.7%. In Figure 1C, lower panel, we report also the absolute number $N_b$ of binders that belong to the largest phase-separated cluster. In general, the number $N_b$ depends both on *c* and $E_{b-b}$. For instance, by varying the affinity $E_{b-b}$ from 2.7 to 4$K_B$T, $N_b/N_{tot}$ ranges from roughly 80% to values higher than 95% ($c \approx 1.9\%$, Figure 1D). For the values of the affinity explored, which fall in the weak biochemical range, and the considered size of the system (see Methods), the minimum number of binders required for the formation of phase-separated clusters results approximately hundreds, which is consistent with the number of molecules (200÷400) found in *in-vivo* stable condensates of PolII and Mediator complexes(5).

All the discussed results are summarized in the phase-diagram in Figure 1E, where the z-axis reports the fraction of system realizations that exhibit phase transition in the time window considered (Methods). As previously specified, the affinity with polymer is $E_{b\text{-}bs} = 3.1 K_B T$. A very similar phase diagram is found for the system made of only binders (Supplementary Figure 1A).

**The interaction with the polymer binding sites can macroscopically influence the system**

Next, we investigate how the presence of the polymer with its binding sites influences the PS process. To this aim, we consider different values of $E_{b\text{-}bs}$ ranging approximately from $3 K_B T$ to $5 K_B T$ (Figure 2A) and study the equilibrium properties of the system by analysing the cluster distribution (see Methods) in each considered condition.

We find that when the binding affinity $E_{b\text{-}b}$ is very low ($\leq 2.5 K_B T$) no macroscopic PS is observed for any value of $E_{b\text{-}bs}$ and of the binder concentration (Supplementary Figure 1B). Nevertheless, for high values of the concentration *c*, small clusters form whose size depends on $E_{b\text{-}bs}$, likely due to the presence of the binding sites that, for high interaction affinities, produce a local increase in the binder density (as sketched in Figure 1A, Supplementary Figure 2A). This effect becomes important and induce the macroscopic transition when the interaction energy $E_{b\text{-}b}$ is increased to values proximal to the transition threshold in the absence of polymer or with a weakly interacting polymer (Methods). Indeed, the presence of the polymer for large $E_{b\text{-}bs}$ triggers the PS transition, for sufficiently high values of *c*, even though the transition is not observed for lower values of $E_{b\text{-}bs}$ and same concentration *c*. This is shown in the bar plots in Figure 2B, where the fraction of binders in the largest cluster is reported for different values of concentration *c* and binding affinity $E_{b\text{-}bs}$.

It is interesting to stress that such macroscopic changes are induced by a very small perturbation to the system, since the concentration of binding sites (around 0.02%, Methods) is roughly two order magnitude lower than the binder concentration range explored. These results suggest that the combined action of the non-specific interaction among the binders and their interaction with the binding sites on the polymer cooperatively influences the evolution of the system.

## The phase-separated cluster exhibits a dynamical structure

In this section we focus on the structural properties of the clusters for different values of the control parameters. To this aim, we observe the behaviour of the cluster by tracking in time, in equilibrium conditions, the distance of its binders from the cluster centre (see Methods), by monitoring the respective trajectories. In Figure 2C, we show, as an example, such distance during time for four different binders, each tracked with a different colour. From these tracks, two fundamental properties emerge. First, the phase-separated cluster is originated from a dynamic equilibrium since binders attach and detach from the cluster surface with a certain frequency (Figure 2C, left panels). Second, the cluster is characterized by a non-trivial internal mobility, with the single binders, adsorbed inside the cluster, moving for distances comparable with the size of the cluster. This last aspect is highlighted by the zoom plots (Figure 2C, right panels). Of course, binders are differently tied to the cluster depending on the interaction affinity $E_{b-b}$, as shown for two values of the interaction affinity reported in the figure ($E_{b-b}$ = 2.7$K_B$T, upper panels and $E_{b-b}$ = 3.0$K_B$T, bottom panels).

More quantitatively, we calculate the fraction of binders $\mu$ that detach from cluster within a fixed time interval (see Methods, Supplementary Figure 2B). Interestingly, for low affinity ($E_{b-b}$ = 2.7$K_B$T), we find that approximately 70% of the binders escape at least one time (Figure 2D) over time scales comparable to the estimated characteristic time between two contact events involving the regulatory elements (see Methods). This implies that such highly dynamic exchange occurs as fast as typical biological processes, in agreement with the liquid-like nature of protein condensates having high exchange rates. Analogously, the estimated average value of attaching time (indicated with $\tau$), i.e. the relative time spent attached to the cluster within a fixed long observation time, results roughly 70% (Figure 2E). Such values vary sensibly upon increase of $E_{b-b}$, e.g. $\mu$ and $\tau$ result about 50% and 90% respectively for $E_{b-b}$ = 3.0$K_B$T. Finally, we focused on the internal mobility of the cluster and calculated the mean-square-displacement (MSD, Figure 2F) of the binders that are always contained in the cluster, by varying the lag time $t_{\text{lag}}$ (see Methods). The behaviour is sub-diffusive since we observe a rough behaviour MSD ~ $(t_{\text{lag}})^\alpha$, with $\alpha$ approximately ranging from 0.5 to 0.7 (see Methods), slightly decreasing with larger $E_{b-b}$.

## Gene-enhancer interaction and cluster dynamics

Next, we investigate how the phase-separated cluster influence the interaction between the binding sites located along the polymer, representing an enhancer and a gene promoter on the chromatin filament.

This is an interesting aspect of the model since in real cells complex structural relationships exist between enhancers and target genes, whose features depend on the genomic region and its regulatory landscape (15). For instance, the contact can be invariant with respect to the transcriptional activity of the gene, as the *Shh* gene with its enhancer *ZRS* (34), or it can be highly tissue-specific, as the *Pitx1* gene and its enhancer *Pen* (35). Many other biological examples are reviewed in reference (15).

In our simplified framework, we can easily study the equilibrium dynamics of the distance between the binding sites, as visually depicted in Figure 3A. We consider three possible scenarios: in the first case (highlighted in green) no phase-separated cluster is formed (that is, low concentration $c$ or low affinity $E_{b-b}$) and the affinity between binders and binding sites ($E_{b-bs}$) is not sufficiently high to mediate stable contacts. Then, the binding sites come in spatial proximity very rarely and interactions only occur as due to random fluctuations of their positions. A typical example of this "free" dynamics is shown in Figure 3B (green curve) where the distance $d_{gene-enh}$ between binding sites is plotted against time. In the second case (highlighted in blue in Figure 3A), the phase-separated cluster mediates the contact between the binding sites with a very strong affinity and completely constraints their motion. The resulting distance dynamics (flat blue curve in Figure 3B) is very stable and practically constant over very long times. Of course, the distance value depends on the relative position of the binding sites when they start to interact with the cluster.

The third case (highlighted in cyan in Figure 3A), occurs when the phase-separated cluster interacts with the binding sites with an intermediate affinity $E_{b-bs}$. In Figure 3C, three examples of distance dynamics are shown which highlight the deep difference with the former cases. Indeed, here the single binding site can detach from the cluster and the distance $d_{gene-enh}$ can increase in short time intervals. This dynamic behaviour is compatible with the dynamic "kissing" model which has been proposed to explain experimental observations where, for instance, the *Esrrb* gene is found to sporadically co-localize with the Mediator cluster (5). Furthermore, the binding sites are much more mobile on the cluster surface. As a consequence, the resulting dynamics, although still influenced by the presence of the cluster, is richer than in the case of strong affinity. By acting on the binder concentration, it is also possible to expand the range of values of the equilibrium distances between the binding sites, as shown in the right-hand plot in Figure 3C (labeled as Case 3), where the higher concentration $c$ ensures a larger equilibrium distance. We stress that the three different dynamic regimes are all observed in equilibrium conditions and naturally emerge by simply varying the control parameters (i.e. $c$, $E_{b-b}$, and $E_{b-bs}$).

The schematic diagram in Figure 3D summarizes the parameters corresponding to each of the described regimes. System details such as number of binding sites used to model the gene or the enhancer (see Methods), that biologically correspond to the regulatory landscape of the region under consideration, can influence the values of concentration and affinity that determine the kind of dynamic regime. Also, more realistic models would require the use of binders with different size, as real proteins span different lengths (approximately in the range 5-30nm (36)). However, we verified that, upon rescaling of the system parameters, similar results are found and the overall dynamical behaviours described above remain qualitatively unchanged (Supplementary Figure 3, Methods).

**Gene-enhancer temporal dynamics**

We compare here our results from the model dynamics with recently published imaging data (33) where the distance between the *Sox2* gene (chr3: 34,548,927-34,551,382, mm9) and its distal super-enhancer *Sox2* Control Region (SCR, chr3: 34653927–34660927, mm9) has been tracked *in-vivo* in different murine tissues. In Figure 3E, we show four examples of such dynamics, each corresponding to a different individual cell. In making the comparison we have considered only embryonic stem (ES) cells, where the *Sox2* gene is active through the contact with SCR (37), as also confirmed by high resolution Hi-C data (38). As discussed in ref. (33), from the experimental tracks it is possible to appreciate the variety of behaviours that individual cells can exhibit, although they all belong to the same type. A first visual comparison with our results shows that the situation corresponding to the cyan dynamical regime in Figure 3A, gives rise to distance dynamics (Figure 3C) that are qualitatively similar to the experimental ones (Figure 3E).

More quantitatively, we computed the histogram of the experimental distances and compared it with the same quantity obtained from our numerical simulations (Figure 3F, Methods). In order to make a meaningful comparison, a suitable scaling of distances is used (Methods). Similarly to the experimental case, we see that the model distribution is not bimodal (33) (Hartigan's Dip test p-val > 0.1 in both cases) and is statistically compatible with the experiments (Kolmogoroff-Smirnov KS test p-val > 0.01). On the contrary, the control distribution obtained from the "free" dynamics case, is not able to describe the experiments (see inset of Figure 3F, KS test p-val = $10^{-24}$). Taken together, these results indicate that the formation of the phase-separated aggregate and its interaction with the gene/enhancer sites act in a cooperative manner to ensure a robust but dynamic contact which can be sometime inhibited by thermal fluctuations.

In this comparison, the chromatin fiber is simply modelled with a uniform string of beads having a fixed genomic content (see Methods). In general, more details could also be taken into account, such as the heterogeneous genomic content along the chromatin, experimentally observed *in-vivo* (39), the above-mentioned size of the binders and a more complex arrangement of binding sites.

**Microscopic mechanisms underlying PS**

In the previous sections we showed that with sufficiently high binding affinity $E_{b\text{-}b}$, the presence of the binding sites along the polymer can induce a macroscopic transition that is not observed for lower affinities (Figure 2B, C). This observation prompted us to investigate the early microscopic dynamics leading to the formation of the phase-separated cluster and the role of the binding sites played in such process.

From our simulations, two main mechanisms emerge, as schematically depicted in Figure 4A. In the first mechanism the PS is not initially influenced by the presence of the binding sites and the cluster mediates the contacts only when it is already (or almost) formed (see Figure 4A, upper panel). In the second mechanism, binding sites induce the formation of small-sized clusters which, acting as nucleation sites, grow and eventually merge together, leading to the full phase transition with the formation of the macroscopic cluster (see Figure 4A, bottom panel). It is worth to mention here that recent technological developments (CasDrop method (40)) allowed to experimentally investigate such process and, in analogy with our findings, have highlighted that artificial protein aggregates, targeted to specific seeded loci, are able to pull in spatial proximity distant chromatin regions.

For a quantitative study of the above mechanisms we have monitored the growth of the largest clusters by tracking their number of binders during time (green curves in Figures 4B and 4C, Methods). In parallel, we have also monitored the average distance between such clusters and the binding sites ($d_{bs\text{-}clust}$, brown curves in Figures 4B and 4C, see Methods). In this way, we tested whether the two processes are connected and therefore which mechanism regulates the system evolution. Note that, in any case, the cluster growth is well described by a linear increase in time, as shown by the black dashed line in Figures 4B and 4C. This is expected from classical results in Statistical Mechanics (41, 42), as we will discuss below. We find that when $E_{b\text{-}bs}$ is comparable or lower than $E_{b\text{-}b}$ (around $3K_BT$ in the considered case), PS is basically unaffected by the presence of the binding sites, as the distance $d_{bs\text{-}clust}$ decays on a time scale (Figure 4B, brown curves) when the cluster is already grown (Figure 4B, green curves). On the other hand, if $E_{b\text{-}bs} \gg E_{b\text{-}b}$, the binder droplets tend to nucleate around the binding sites, as witnessed by the fact that the decay time of $d_{bs\text{-}clust}$ is approximately one order of magnitude shorter than in the previous case (brown curves, Figure 4C, Methods). In this case the contact between binding sites occurs when the cluster counts just few binders (approximately tens of binders, Figure 4C, green curves). The results obtained with different model parameters are summarized in Figure 4D, where the average distance between the binding sites and the major clusters at early times, well before PS has been completed (Methods), is reported for different values of $c$ and $E_{b\text{-}bs}$, $E_{b\text{-}b}$ being $\sim 3K_BT$. The described analysis, where cluster growth and distance between binding sites and clusters are observed during time, could be a possible experimental strategy to understand the dynamics mechanism bridging distant chromatin regions *in-vivo*.

Finally, we have investigated the microscopic process driving the formation of the phase-separated clusters. To this aim, we performed a highly time resolved analysis (Methods) of the growth of the cluster. In Figure 4E, two independent examples (labelled as 1 and 2) are reported, where the number of binders $N_b$ belonging to the major four clusters is tracked in time (Methods). From these plots, it is possible to appreciate two different mechanisms occurring at microscopic level. In the upper plot (Dynamics 1 in Figure 4E), the first and second clusters (red and green curves) regularly grow and compete, until the red dominates and the green continuously decrease and eventually drops to negligible values. This behaviour is the signature of an evaporation-condensation process, where the green cluster undergoes a gradual loss of its binders that are adsorbed by the red one after travelling between the two droplets. Conversely, in the bottom plot (Dynamics 2 in Figure 4E), the green curve suddenly disappears while the red curve has a discontinuous jump indicating that the two clusters merged together in a coalescence event. Alternatively, it is possible to appreciate the different processes by considering the entire distribution of the cluster size during time, which is dynamically shown in Supplementary Movie 1 and 2. Evaporation-condensation and coalescence (schematically shown in Figure 4F) are therefore the two fundamental microscopic mechanisms that compete / cooperate to the overall PS process. From the theoretical point of view, it is well established (41–45) that in a three-dimensional system both mechanisms lead to the linear increase of the volume of the aggregate, as indeed confirmed by our simulations (Figures 4B, C).

Biologically, coalescence is a well-known process for phase-separated protein aggregates and has been observed in several studies (2, 3, 5, 40). On the other hand, transient small clusters have also been observed (e.g. PolII and Mediator) (5) and could be identified with the transient clusters shrinking because of evaporation. By experimental implementation of the described approach, e.g. by monitoring in time size and geometry of the clusters (as suggested in (31)) with live imaging techniques at high temporal resolution (5), it would be possible to understand how the two processes are related in real cells and the conditions in which they occur.

**CONCLUSIONS**

In this work, we investigated the biophysical mechanisms behind the PS process that regulates the interaction between genes and enhancers, by using an essential polymer model with binding sites simulating genomic regulatory elements and non-specifically interacting multivalent binders modelling the proteins, such as TFs, that are present in the nuclear environment and are known to interact with chromatin. Typically, implementations of this polymer model, more commonly known as the Strings and Binders Switch (29) (SBS) or the TFs (27) model, do not take into account the weak non-specific interaction among molecules, even though the formation of clusters spontaneously emerge as bridging-induced micro phase-separation (30).

We showed that such model exhibits phase-transitions, leading to the formation of particle clusters, when the binding affinity among the molecules and molecular concentration are above threshold values. The structural properties of the molecular aggregates, as number of molecules, mobility and exchange rate, depend on the system parameters and, upon certain conditions, can be highly dynamic in agreement with the liquid-like nature of protein condensates, experimentally highlighted by photobleaching experiments (3, 30). By varying the interaction affinity between the binders and the binding sites of the polymer, that are genes or enhancers, it is possible to obtain gene-enhancer contact dynamics similar to live-imaging experimental data (33) and compatible with the "kissing" model proposed to explain co-localization data of protein clusters with the *Esrrb* gene (5). Conversely, the presence of the polymer with its binding sites can influence the PS process, since it can be induced and catalyzed by a nucleation mechanism that can produce macroscopic effects. The dynamic and equilibrium properties of the system are therefore the result of a collective behaviour emerging from the interplay between PS and interactions among the binding molecules, genes and enhancers.

Naturally, the model can be generalized in different ways, e.g. by introducing more types of binders with different sizes and specific interactions, which have been shown to generate multi-layered aggregates experimentally observed (11), as well as more details accounting for the protein structure, as the DNA Binding Domains (DBDs) and the Intrinsically Disordered Regions (IDRs) (6). Also, more complex arrangements of binding sites along the polymer, better modelling the regulatory landscape of real genomic regions, can be employed in order to investigate more accurately the associated chromatin architecture (23, 24, 46). Furthermore, the presence of a mechano-active surrounding environment which influences and, in turn, is influenced by the formation of phase-separated cluster (40), is neglected in our study and could be implemented, e.g., with a dense viscoelastic matrix of polymers (47). Nevertheless, although based on an essential model, our work provides a theoretical framework able to recapitulate many features of PS mechanism and its interplay with the gene-enhancer communication.

# MAIN FIGURES

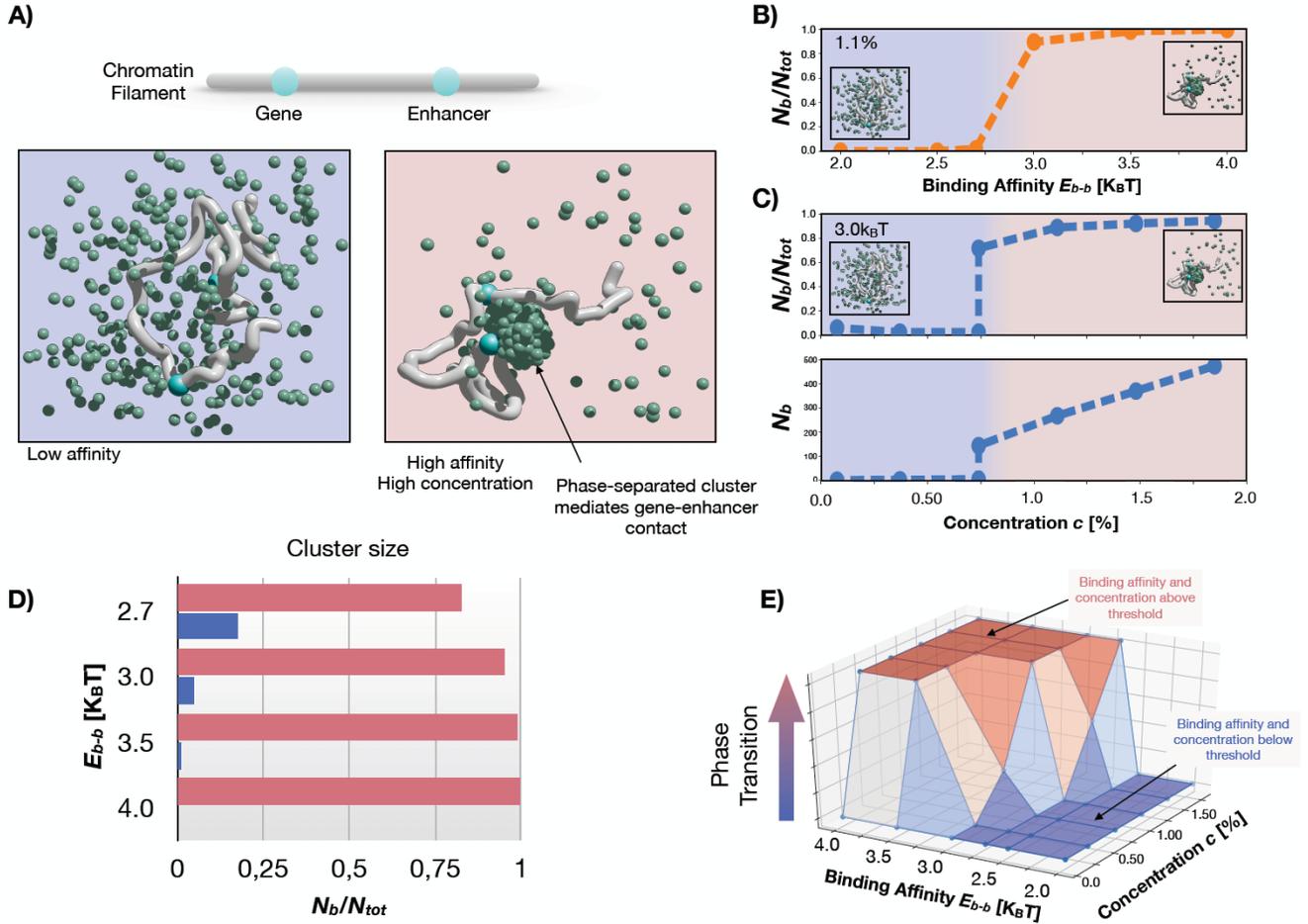

**Figure 1: Phase-separation (PS) is a switch-like process mediating the formation of the gene-enhancer interaction**

**A)** The formation of a phase-separated cluster can mediate the contact between a gene and its enhancer. **B)** The fraction of binders $N_b/N_{tot}$ in the phase-separated cluster as a function of the binding affinity $E_{b-b}$ among the binders at equilibrium. The concentration is $c \sim 1.1\%$. The transition occurs between $2.7K_BT$ and $3.0K_BT$. **C)** Fraction of binders in the cluster as a function of the binder concentration $c$. The binding affinity is $E_{b-b} = 3.0K_BT$. At the transition point $c \sim 0.7\%$, in some cases the phase-separated could not form, within the time window considered. **D)** The horizontal red (blue) bar shows the fraction of binders belonging (not belonging) to the cluster for different binding affinities $E_{b-b}$ (y-axis) at high concentration ($c \sim 1.9\%$). **E)** Three-dimensional phase-diagram of the system. On the z-axis it is reported the fraction of times in which the system exhibited the phase transition for a fixed parameter configuration. For all panels, interaction with binding sites $E_{b-bs}$ is $3.1K_BT$. Similar results are obtained in the absence of polymer.

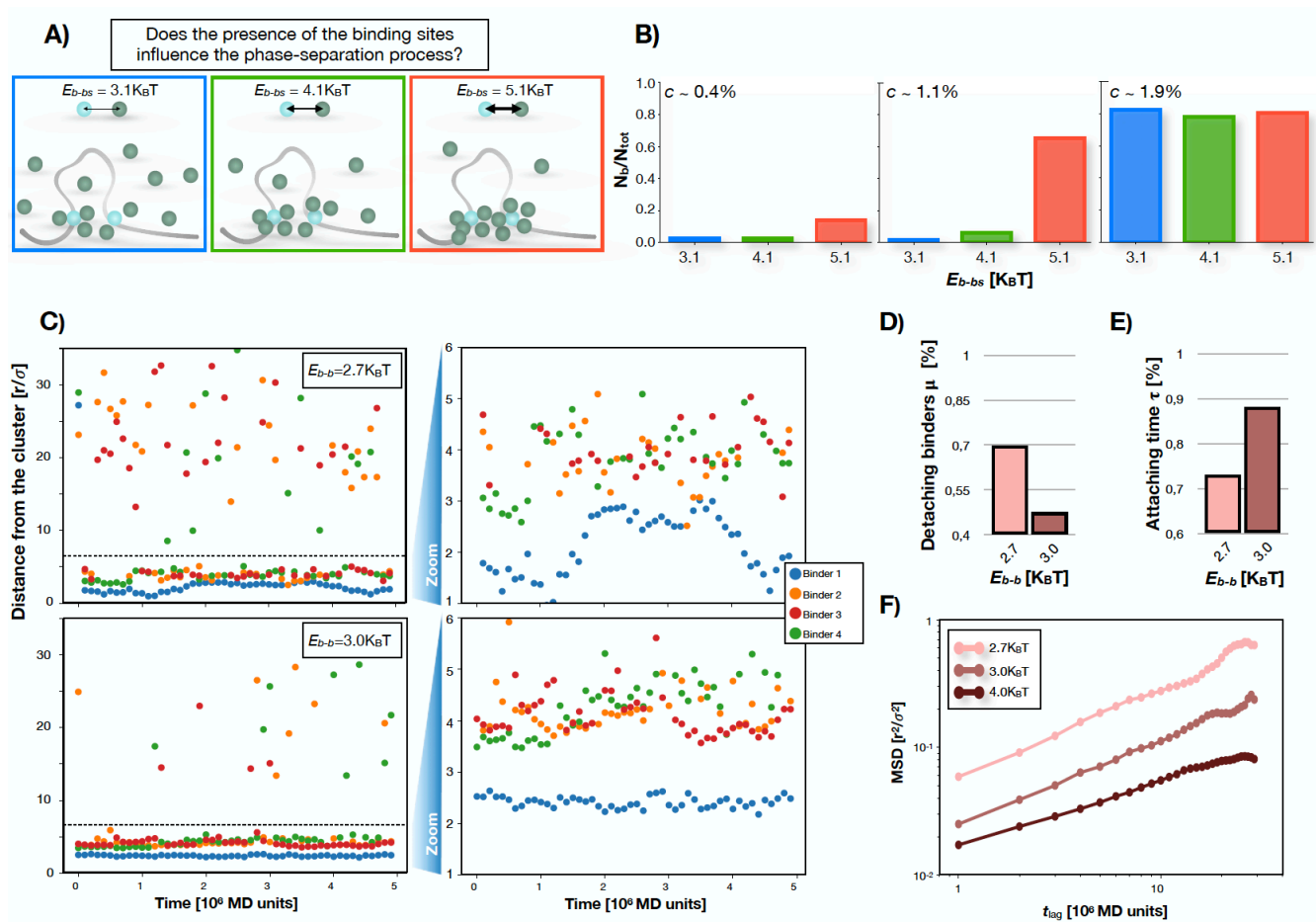

**Figure 2: Influence of the binding sites and structural properties of the phase-separated clusters**

**A)** The interaction between the binding sites and the binders influences the system properties by locally increasing the concentration around the binding sites. **B)** Fraction of total binders forming the largest cluster, $N_b/N_{tot}$, for different concentrations $c$ and $E_{b-bs}$, at the transition energy $E_{b-b} = 2.7 K_B T$. For $c \sim 1.1\%$, only at $E_{b-bs} = 5.1 K_B T$ the system exhibits a macroscopic transition. **C)** Left panels: distance between single binders and the centre of the phase-separated cluster, as function of time, for two different values of $E_{b-b}$. Each colour represents a different binder. The horizontal dashed line indicates an estimate of the cluster size. When the dot is above the line, the binder is escaped from the cluster. Right panels: zoom highlighting the mobility of the binders when they move within the cluster. **D)** Fraction of binders $\mu$ that detach from the cluster at least once in a characteristic time interval. For low affinities, the cluster exhibits highly dynamical properties since up to roughly 70% of the binders experience a detaching event. **E)** The attaching time $\tau$, that is the average relative time spent by a single binder in the cluster, strongly depends on the binding affinity $E_{b\text{-}b}$. **F)** The Mean Square Displacement (MSD, log-log scale) of the binders within the cluster, for different values of $E_{b\text{-}b}$. As the energy decreases, the binders have a higher mobility and the cluster becomes more dynamic. For panels C, D, E and F the concentration used is $c \sim 1.5\%$.

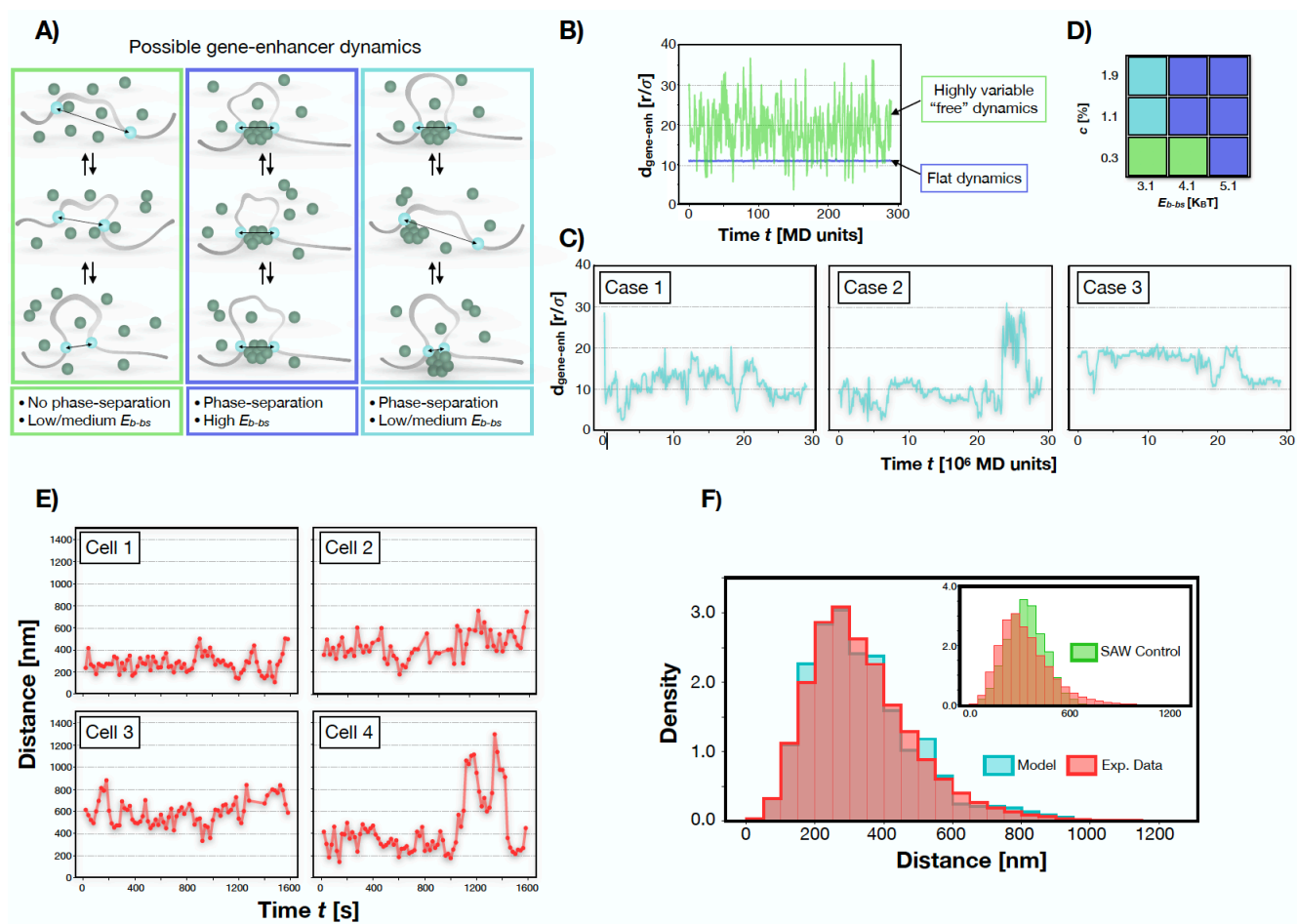

**Figure 3: The PS process produces different regimes of gene-enhancer contact dynamics**

**A)** The affinity between binding sites and binders influence the gene-enhancer contacts in time and different dynamics emerge. **B)** Example of gene-enhancer distance dynamics where a stable contact is not formed (green curve, $E_{b\text{-}bs} = 3.1 K_B T$, $E_{b\text{-}b} = 3.0 K_B T$ and $c \sim 0.3\%$ and no phase-separated cluster) and where a highly stable contact (blue curve, $E_{b\text{-}bs} = 5.1 K_B T$, $E_{b\text{-}b} = 3.0 K_B T$ and $c \sim 1.1\%$) mediated by a phase-separated cluster is formed. **C)** Three examples of contact dynamics for intermediate affinity $E_{b-bs}$ = 3.1K$_B$T ($E_{b-b}$ = 3.0K$_B$T). Case 1 and 2 are with $c$ ~ 1.1% and case 3 with $c$ ~ 1.9%. Here the contact is stable, yet it results much more variable as the binding sites can move on the surface of the cluster and can also detach. **D)** Diagram summarizing the different dynamic regimes and their corresponding parameters. **E)** Examples of experimental single cell gene-enhancer dynamics from the *Sox2* and its super-enhancer SCR, in mouse ES cells. Data taken from ref. (33). The profile results very similar to the simulated dynamics in panel C. **F)** Comparison between the model (cyan) and experimental (red) distributions of distances. The distributions are consistent (Kolmogoroff-Smirnov KS test p-val > 0.01). The inset shows the comparison with the control distribution of the free SAW case (KS test p-val = 10$^{-24}$).

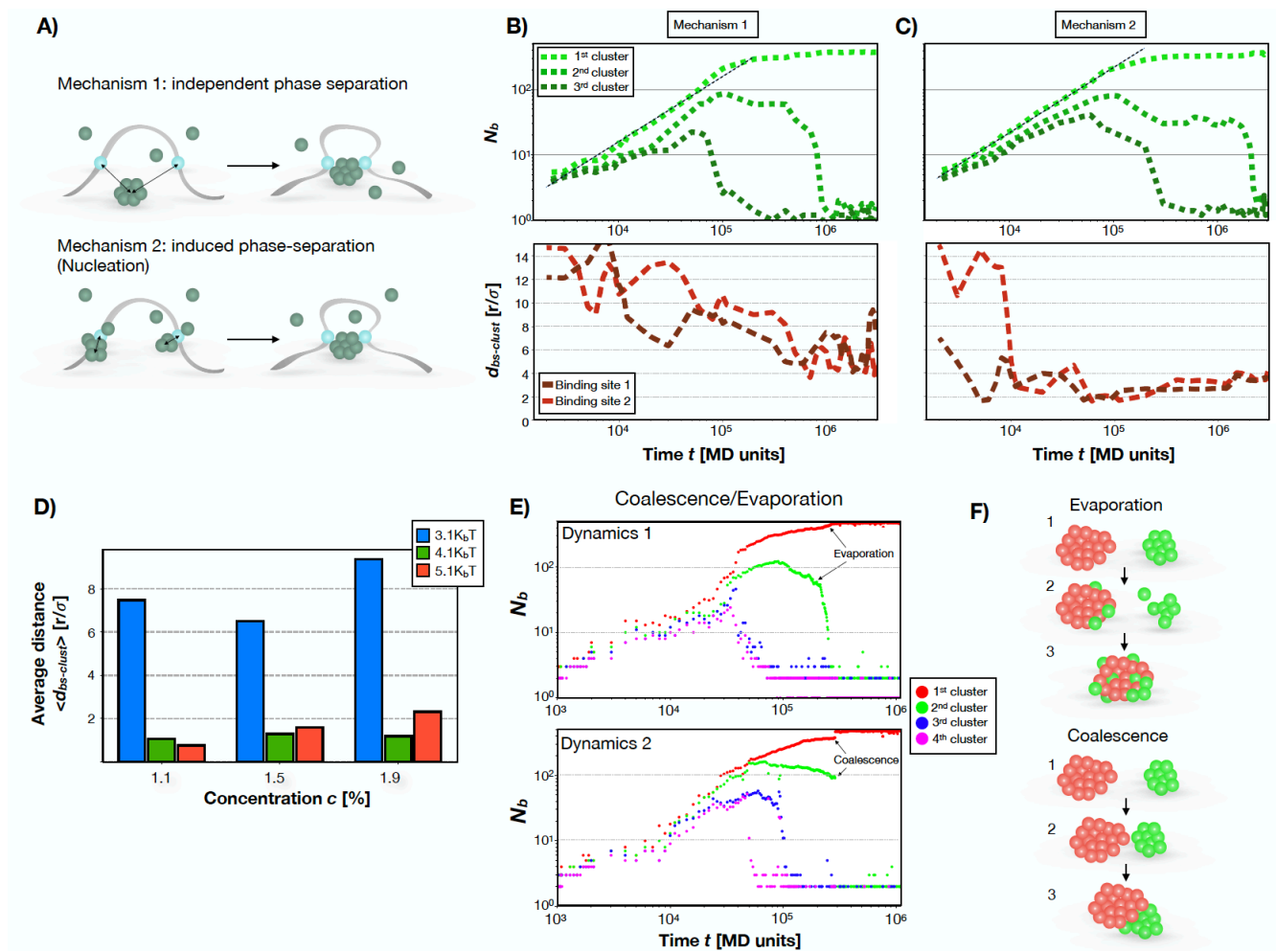

**Figure 4: Microscopic mechanisms of PS induction**

**A)** Possible mechanisms leading to PS and driving contact formation between regulatory elements. **B)**

Number of binders $N_b$ composing the three main clusters (green curves). Note the linear increase during time. Below, distance of the main clusters from the binding sites ($d_{bs\text{-}clust}$, brown curves). Here, binding affinity $E_{b\text{-}bs}$ = 3.1$K_B$T, concentration $c \sim 1.5\%$. **C)** As B, with $E_{b\text{-}bs}$ = 5.1$K_B$T. **D)** The average distance $<d_{bs\text{-}clust}>$ at early stage of the dynamics (Methods) for different concentrations $c$ and binding affinities $E_{b\text{-}bs}$. **E)** The number of binders $N_b$ of the first four largest clusters monitored over a highly time resolved simulation. In Dynamics 1, the gradual and continuous decrease of the top second cluster (green curve) and the gradual increase of the top first cluster (red curve) indicate an evaporation process. Conversely, in Dynamics 2 the top second cluster drops and the top first cluster increases respectively in a discontinuous way, indicating the coalescence between the two (green+red). Simulations were performed with $E_{b\text{-}b}$ = 3.0$K_B$T, $E_{b\text{-}bs}$ = 3.1$K_B$T and $c \sim 1.9\%$. **F)** Schematic representation of evaporation and coalescence, the microscopic mechanisms that regulate PS dynamics.

## ACKNOWLEDGEMENTS

We acknowledge funding MIUR PRIN 2015K7KK8 "Statistical Mechanics and Complexity". We thank Maria Barbi for interesting discussion and Simona Bianco for critical read and helpful remarks. A.M.C acknowledges computer resources from ENEA CRESCO/ENEAGRID (48).
The authors declare no conflicts of interest.

## AUTHOR CONTRIBUTIONS

A.M.C, F.C. and M.S. designed the research project. A.M.C. run computer simulations and performed data analyses. A.M.C, F.C. and M.S. wrote the manuscript.

## METHODS

<u>Model details</u>

To investigate the interplay between PS process and gene-enhancer dynamic, we use a simple polymer model, based on the Strings and Binders Switch (SBS) polymer model (29) (also known as TFs model (27)), in which a chromatin filament is modelled by a Self-Avoiding-Walk (SAW) string made of $N$ beads, with some binding sites that can interact with binding factors (or simply binders) floating in the surrounding environment. Although the model envisages the possibility of different types of binders (25), the results discussed here are based, for sake of simplicity, on models with only one type of binder. The binders can interact with binding sites placed on the polymer, with an interaction affinity $E_{b\text{-}bs}$ and a total concentration *c*. Furthermore, binders interact among themselves with an attractive multivalent interaction $E_{b\text{-}b}$. In all simulations, we use polymers with $N = 200$. The total number of binding sites is 6, arranged in two groups of 3 sites symmetrically located on the polymer with an average linear distance of 100 beads. To check the robustness of the results, we used also polymers with arrangements having 4 and 2 binding sites and found, in general, similar behaviours.

Molecular Dynamics simulations details

The entire system (polymer beads and binders) is subject to thermal fluctuations at temperature T, so the particles obey to the Langevin equation (49). Unless differently stated, beads and binders have same diameter $\sigma$ and mass $m$ that we set equal to 1, in dimensionless units (50). We use a purely repulsive Lennard-Jones (LJ) potential between any two particles in order to account for excluded volume effects, with length scale $\sigma$ and energy scale $\varepsilon$ measured in $K_B T$ units (50). Between any two adjacent beads of the polymer we use a finitely extensible non-linear elastic spring (FENE (50)), with standard parameters (27, 51, 52) (length constant $R_0 = 1.6\sigma$ and spring constant $K = 30 K_B T/\sigma^2$).

The interaction between beads that are binding sites representing genes or enhancers and binders as well as the non-specific interaction among the binders was modelled by a short-range, truncated attractive LJ potential $V_{LJ}$ in the form:

$$V_{LJ}(r) = 4\varepsilon \left[ \left(\frac{\sigma}{r}\right)^{12} - \left(\frac{\sigma}{r}\right)^{6} - \left(\frac{\sigma}{R_{int}}\right)^{12} + \left(\frac{\sigma}{R_{int}}\right)^{6} \right]$$

for $r < R_{int} = 1.3\sigma$, 0 otherwise. The interaction affinities reported in the figures (named $E_{b\text{-}b}$ and $E_{b\text{-}bs}$) are given by the minimum of $V_{LJ}$ and are controlled by $\varepsilon$ (53). In our simulations, $\varepsilon$ was sampled in the range (5.8÷11.6) $K_B T$ for $E_{b\text{-}b}$ and in the range (9÷15) $K_B T$ for $E_{b\text{-}bs}$.

The Langevin equation is integrated using the online available LAMMPS package(54). The dynamics parameters are set to standard values(53), that is friction coefficient $\zeta = 0.5$, temperature T = 1 and integration time step dt = 0.012 (50, 55), expressed in dimensionless units. The system is confined in a cubic simulation box with periodic boundary conditions, with edge size D = $30\sigma$ for the simulations in Figures 1, 2 and 4 and D = $60\sigma$, in order to minimize finite size effects, for the simulations presented in Figure 3.

Each polymer is initialized to a random Self-Avoiding-Walk (SAW) polymer configuration (50). The binders are uniformly distributed in the box as the simulation starts, with a concentration per volume unit $c = (4\pi r^3/3)*N_{tot}/D^3$, where $N_{tot}$ is the total number of binders and $r$ the radius of the binder. Note that $c$ is linked to the molar concentration $c_m$ through the relation $c = (c_m\sigma^3)*N_A$, where $N_A$ is the Avogadro number and $\sigma$ is the physical length scale. In the figures, concentrations are rescaled by the geometric factor $4\pi r^3/3 \approx 0.5$. To check the robustness of the results, we verified that a reduction of the binder size led to analogous results upon rescaling of other quantities (as $N_{tot}$ and D) so to keep the binder concentration per volume unit $c$ in the explored range of values.

For each parameter choice, we perform 10 independent simulations, which were equilibrated up to $20*10^7$ timesteps, so to ensure the phase-transition. Starting from the initial state, the configurations were sampled logarithmically in time, and then, after $10^5$ timesteps, taken every $10^5$ timesteps, except for simulations presented in Figures 4E and Supplementary Movies S1 and S2, where configurations were sampled every $2*10^3$ timesteps.

Clustering method

To obtain size distributions of binder aggregates, we use a standard iterative clustering method where the Euclidean distance among the binders is used as metric. In order to define the clusters, we set as threshold distance the cutoff $R_{int}$ of the LJ interaction and define as a cluster the set of particles whose mutual distance is smaller than $R_{int}$, using the Python package "*scipy.cluster*". To avoid wrong counts in the cluster number, for each frame analyzed we implemented iterative boundary corrections, where the simulation box is replicated at each side and the configuration having the minimum number of clusters is considered. At equilibrium, when the fraction of binders $N_b/N_{tot}$ in the main cluster is above 50%, then a PS event is considered. In the phase-diagram of Figure 1E, the fraction of system realizations satisfying this condition is reported. When the system is close to the critical threshold, the time required to observe the full PS is longer. In these cases, some replicates do not exhibit the transition in the time window considered. Here, the fraction of realizations where the transition occurs is less than 1.

Gene-enhancer distance dynamics

Distances $d_{gene-enh}$ are extracted from simulated 3D trajectories in equilibrium conditions, that is when the phase-separated cluster is formed. For each control parameter combination, distance distributions are obtained by considering ten independent equilibrium trajectories. For sake of simplicity, each polymer binding regions is made by 3 binding sites and have an average separation of 100 beads.

Nevertheless, we verified that by varying separation and number of binding sites, analogous dynamics regimes are found, as shown in Supplementary Figure 3, where binding regions are made by 1 and 3 binding sites respectively, separated by 50 beads.

Mixtures of systems using different concentrations are used for the comparison with experimental data. In Figure 3F a 0.25:0.75 mixture with $c \sim 1.9\%$ and $c \sim 1.1\%$ respectively is shown. By equating the experimental and model averages (24, 46), we obtain the factor $\sigma \approx 29$nm that maps the dimensionless length scale in physical units. Since the genomic distance between Sox2 and SCR is ~100kb, the genomic content of each bead in our polymer results ~1kb. Note that, in this case, equilibrium clusters count thousands of binders since simulations were performed with a bigger box to avoid finite size effects. Experimental trajectories are taken from reference (33), in mouse embryonic stem (ES) cells.

As a characteristic biological time to compare with, we consider the average time between two contact events, defined when $d_{gene-enh}$ is shorter than the cluster diameter. For the parameters range explored, values of characteristic time roughly range in the interval $3 \div 4 * 10^6$ timesteps.

Equilibrium properties of the phase-separated cluster

The fraction of binders $\mu$ that detaches from the cluster is defined as $(N_b - N_{in})/N_b$, where $N_b$ is the average number of binders in the cluster and $N_{in}$ is the number of binders that never escape from the cluster, in a fixed time interval. A binder is considered detached if its distance from the cluster center is larger than the estimated size of the cluster. Analogously, the average attaching time $\tau$ spent by a single binder in the cluster in a time interval $\Delta t$ is defined as $<t_{in}/\Delta t>$, where, $t_{in}$ is the number of times that the binder is found attached to the cluster and $<>$ indicates average over the binders. In Figure 2D and 2E, the reported values are computed in a time interval $\Delta t = 4.5*10^6$ timesteps, in equilibrium conditions, where they result in a plateau regime, as shown in Supplementary Figure 2B.

The Mean Square Displacement (MSD) is calculated according to the standard formula $\text{MSD}(t_{\text{lag}}) = \sum_{t=0}^{\Delta t - t_{\text{lag}}} \frac{1}{(\Delta t - t_{\text{lag}})} (d(t+t_{\text{lag}}) - d(t))^2$, where $d(t)$ is the distance from the center of the cluster at time $t$ and $\Delta t$ is the time interval considered. MSD is calculated and averaged over the binders that never escape from the cluster in the time interval $\Delta t$. A power-law fit $\text{MSD}(t_{\text{lag}}) = a*(t_{\text{lag}})^\alpha$ is performed over the first ten $t_{\text{lag}}$ points, as for higher values of $t_{\text{lag}}$ the number of configurations in the sum reduces and the fluctuations of MSD are larger. For all the described quantities, ensemble averages are performed.

Dynamics of the cluster formation

To study the dynamics of the cluster formation, we consider system configurations from the initial state up to equilibrium states. For each configuration, we find the distribution of the cluster size $N_b$ and select the largest ones. At each time, an ensemble average is performed, namely over the different replicas of the dynamical process.

The distance $d_{bs\text{-}clust}$ between one specific polymer binding site (bs) and the clusters, plotted in Figure 4B and 4C, is defined as $\min_i([dist(bs, clust_i)])$, where "i" labels the first three major clusters, ranked according their size $N_b$. Again, an ensemble average is performed. Estimation of the decay time for $d_{bs\text{-}clust}$ is made with an exponential fit $d_{bs\text{-}clust}(t) = a*\exp(t/b)+c$ and returns b ≈ $1.3*10^5$ and b ≈ $7.1*10^3$, expressed in MD timesteps, for $E_{b\text{-}bs} = 3.1 K_B T$ and $E_{b\text{-}bs} = 5.1 K_B T$ respectively.

Average $<d_{bs\text{-}clust}>$ in Figure 4D is defined as $\frac{1}{3}\sum_{t=1}^{3} \min_j(\min_i([dist(bs_j, clust_i)]))$, where the sum runs over the first three timesteps with at least one cluster with $N_b > 15$, "i" labels the first three major clusters and "j" labels the two binding sites.

**REFERENCES**

1. Banani, S.F., H.O. Lee, A.A. Hyman, and M.K. Rosen. 2017. Biomolecular condensates: organizers of cellular biochemistry. *Nat. Rev. Mol. Cell Biol.* 18:285–298.
2. Shin, Y., and C.P. Brangwynne. 2017. Liquid phase condensation in cell physiology and disease. *Science.* 357:eaaf4382.
3. Sabari, B.R., A. Dall'Agnese, A. Boija, I.A. Klein, E.L. Coffey, K. Shrinivas, B.J. Abraham, N.M. Hannett, A. V. Zamudio, J.C. Manteiga, C.H. Li, Y.E. Guo, D.S. Day, J. Schuijers, E. Vasile, S. Malik, D. Hnisz, T.I. Lee, I.I. Cisse, R.G. Roeder, P.A. Sharp, A.K. Chakraborty, and R.A. Young. 2018. Coactivator condensation at super-enhancers links phase separation and gene control. *Science.* 361:eaar3958.
4. Cho, W.-K., N. Jayanth, B.P. English, T. Inoue, J.O. Andrews, W. Conway, J.B. Grimm, J.-H. Spille, L.D. Lavis, T. Lionnet, and I.I. Cisse. 2016. RNA Polymerase II cluster dynamics predict mRNA output in living cells. *Elife*. 5:e13617.
5. Cho, W.-K., J.-H. Spille, M. Hecht, C. Lee, C. Li, V. Grube, and I.I. Cisse. 2018. Mediator and RNA polymerase II clusters associate in transcription-dependent condensates. *Science.* 361:412–415.
6. Boija, A., I.A. Klein, B.R. Sabari, A. Dall'Agnese, E.L. Coffey, A. V. Zamudio, C.H. Li, K. Shrinivas, J.C. Manteiga, N.M. Hannett, B.J. Abraham, L.K. Afeyan, Y.E. Guo, J.K. Rimel,

C.B. Fant, J. Schuijers, T.I. Lee, D.J. Taatjes, and R.A. Young. 2018. Transcription Factors Activate Genes through the Phase-Separation Capacity of Their Activation Domains. *Cell*. 175:1842-1855.e16.

7. Hnisz, D., K. Shrinivas, R.A. Young, A.K. Chakraborty, and P.A. Sharp. 2017. A Phase Separation Model for Transcriptional Control. *Cell*. 169:13–23.

8. Strom, A.R., A. V. Emelyanov, M. Mir, D. V. Fyodorov, X. Darzacq, and G.H. Karpen. 2017. Phase separation drives heterochromatin domain formation. *Nature*. 547:241–245.

9. Larson, A.G., D. Elnatan, M.M. Keenen, M.J. Trnka, J.B. Johnston, A.L. Burlingame, D.A. Agard, S. Redding, and G.J. Narlikar. 2017. Liquid droplet formation by HP1α suggests a role for phase separation in heterochromatin. *Nature*. 547:236–240.

10. Gibson, B.A., L.K. Doolittle, M.W.G. Schneider, L.E. Jensen, N. Gamarra, L. Henry, D.W. Gerlich, S. Redding, and M.K. Rosen. 2019. Organization of Chromatin by Intrinsic and Regulated Phase Separation. *Cell*. 179:470-484.e21.

11. Boeynaems, S., A.S. Holehouse, V. Weinhardt, D. Kovacs, J. Van Lindt, C. Larabell, L. Van Den Bosch, R. Das, P.S. Tompa, R. V. Pappu, and A.D. Gitler. 2019. Spontaneous driving forces give rise to protein−RNA condensates with coexisting phases and complex material properties. *Proc. Natl. Acad. Sci.* 116:7889–7898.

12. Barbi, M., C. Place, V. Popkov, and M. Salerno. 2004. A Model of Sequence-Dependent Protein Diffusion Along DNA. *J. Biol. Phys.* 30:203–226.

13. Dekker, J., and L. Mirny. 2016. The 3D Genome as Moderator of Chromosomal Communication. *Cell*. 164:1110–1121.

14. Dixon, J.R., D.U. Gorkin, and B. Ren. 2016. Chromatin Domains: The Unit of Chromosome Organization. *Mol. Cell*. 62:668–680.

15. Robson, M.I., A.R. Ringel, and S. Mundlos. 2019. Regulatory Landscaping: How Enhancer-Promoter Communication Is Sculpted in 3D. *Mol. Cell*. 74:1110–1122.

16. Spielmann, M., D.G. Lupiáñez, and S. Mundlos. 2018. Structural variation in the 3D genome. *Nat. Rev. Genet.* 19:453–467.

17. Fiorillo, L., S. Bianco, A. Esposito, M. Conte, R. Sciarretta, F. Musella, and A.M. Chiariello. 2020. A modern challenge of polymer physics: Novel ways to study, interpret, and reconstruct chromatin structure. *WIREs Comput. Mol. Sci.* 10.

18. Bianco, S., A.M. Chiariello, M. Conte, A. Esposito, L. Fiorillo, F. Musella, and M. Nicodemi. 2020. Computational approaches from polymer physics to investigate chromatin folding. *Curr.*

*Opin. Cell Biol.* 64:10–17.

19. Portillo-Ledesma, S., and T. Schlick. 2020. Bridging chromatin structure and function over a range of experimental spatial and temporal scales by molecular modeling. *WIREs Comput. Mol. Sci.* 10.

20. Sanborn, A.L., S.S.P. Rao, S.-C. Huang, N.C. Durand, M.H. Huntley, A.I. Jewett, I.D. Bochkov, D. Chinnappan, A. Cutkosky, J. Li, K.P. Geeting, A. Gnirke, A. Melnikov, D. McKenna, E.K. Stamenova, E.S. Lander, and E.L. Aiden. 2015. Chromatin extrusion explains key features of loop and domain formation in wild-type and engineered genomes. *Proc. Natl. Acad. Sci.* 112:E6456–E6465.

21. Fudenberg, G., M. Imakaev, C. Lu, A. Goloborodko, N. Abdennur, and L.A. Mirny. 2016. Formation of Chromosomal Domains by Loop Extrusion. *Cell Rep.* 15:2038–2049.

22. Buckle, A., C.A. Brackley, S. Boyle, D. Marenduzzo, and N. Gilbert. 2018. Polymer Simulations of Heteromorphic Chromatin Predict the 3D Folding of Complex Genomic Loci. *Mol. Cell.* 72:786-797.e11.

23. Bianco, S., C. Annunziatella, G. Andrey, A.M. Chiariello, A. Esposito, L. Fiorillo, A. Prisco, M. Conte, R. Campanile, and M. Nicodemi. 2019. Modeling Single-Molecule Conformations of the HoxD Region in Mouse Embryonic Stem and Cortical Neuronal Cells. *Cell Rep.* 28:1574-1583.e4.

24. Chiariello, A.M., S. Bianco, A.M. Oudelaar, A. Esposito, C. Annunziatella, L. Fiorillo, M. Conte, A. Corrado, A. Prisco, M.S.C. Larke, J.M. Telenius, R. Sciarretta, F. Musella, V.J. Buckle, D.R. Higgs, J.R. Hughes, and M. Nicodemi. 2020. A Dynamic Folded Hairpin Conformation Is Associated with α-Globin Activation in Erythroid Cells. *Cell Rep.* 30:2125-2135.e5.

25. Bianco, S., D.G. Lupiáñez, A.M. Chiariello, C. Annunziatella, K. Kraft, R. Schöpflin, L. Wittler, G. Andrey, M. Vingron, A. Pombo, S. Mundlos, and M. Nicodemi. 2018. Polymer physics predicts the effects of structural variants on chromatin architecture. *Nat. Genet.* 50:662–667.

26. Bascom, G.D., C.G. Myers, and T. Schlick. 2019. Mesoscale modeling reveals formation of an epigenetically driven HOXC gene hub. *Proc. Natl. Acad. Sci.* 116:4955–4962.

27. Brackley, C.A., S. Taylor, A. Papantonis, P.R. Cook, and D. Marenduzzo. 2013. Nonspecific bridging-induced attraction drives clustering of DNA-binding proteins and genome organization. *Proc. Natl. Acad. Sci.* 110:E3605–E3611.

28. Chiariello, A.M., C. Annunziatella, S. Bianco, A. Esposito, and M. Nicodemi. 2016. Polymer

physics of chromosome large-scale 3D organisation. *Sci. Rep.* 6:29775.

29. Barbieri, M., M. Chotalia, J. Fraser, L.-M. Lavitas, J. Dostie, A. Pombo, and M. Nicodemi. 2012. Complexity of chromatin folding is captured by the strings and binders switch model. *Proc. Natl. Acad. Sci.* 109:16173–16178.

30. Brackley, C.A., and D. Marenduzzo. 2020. Bridging-induced microphase separation: photobleaching experiments, chromatin domains and the need for active reactions. *Brief. Funct. Genomics*. 19:111–118.

31. Erdel, F., and K. Rippe. 2018. Formation of Chromatin Subcompartments by Phase Separation. *Biophys. J.* 114:2262–2270.

32. Scialdone, A., I. Cataudella, M. Barbieri, A. Prisco, and M. Nicodemi. 2011. Conformation Regulation of the X Chromosome Inactivation Center: A Model. *PLoS Comput. Biol.* 7:e1002229.

33. Alexander, J.M., J. Guan, B. Li, L. Maliskova, M. Song, Y. Shen, B. Huang, S. Lomvardas, and O.D. Weiner. 2019. Live-cell imaging reveals enhancer-dependent Sox2 transcription in the absence of enhancer proximity. *Elife*. 8:e41769.

34. Paliou, C., P. Guckelberger, R. Schöpflin, V. Heinrich, A. Esposito, A.M. Chiariello, S. Bianco, C. Annunziatella, J. Helmuth, S. Haas, I. Jerković, N. Brieske, L. Wittler, B. Timmermann, M. Nicodemi, M. Vingron, S. Mundlos, and G. Andrey. 2019. Preformed chromatin topology assists transcriptional robustness of Shh during limb development. *Proc. Natl. Acad. Sci.* 116:12390–12399.

35. Kragesteen, B.K., M. Spielmann, C. Paliou, V. Heinrich, R. Schöpflin, A. Esposito, C. Annunziatella, S. Bianco, A.M. Chiariello, I. Jerković, I. Harabula, P. Guckelberger, M. Pechstein, L. Wittler, W.L. Chan, M. Franke, D.G. Lupiáñez, K. Kraft, B. Timmermann, M. Vingron, A. Visel, M. Nicodemi, S. Mundlos, and G. Andrey. 2018. Dynamic 3D chromatin architecture contributes to enhancer specificity and limb morphogenesis. *Nat. Genet.* 50:1463–1473.

36. Maeshima, K., K. Kaizu, S. Tamura, T. Nozaki, T. Kokubo, and K. Takahashi. 2015. The physical size of transcription factors is key to transcriptional regulation in chromatin domains. *J. Phys. Condens. Matter*. 27:064116.

37. Li, Y., C.M. Rivera, H. Ishii, F. Jin, S. Selvaraj, A.Y. Lee, J.R. Dixon, and B. Ren. 2014. CRISPR Reveals a Distal Super-Enhancer Required for Sox2 Expression in Mouse Embryonic Stem Cells. *PLoS One*. 9:e114485.

38. Bonev, B., N. Mendelson Cohen, Q. Szabo, L. Fritsch, G.L. Papadopoulos, Y. Lubling, X. Xu, X. Lv, J.P. Hugnot, A. Tanay, and G. Cavalli. 2017. Multiscale 3D Genome Rewiring during Mouse Neural Development. *Cell*. 171:557-572.e24.

39. Ricci, M.A., C. Manzo, M.F. García-Parajo, M. Lakadamyali, and M.P. Cosma. 2015. Chromatin Fibers Are Formed by Heterogeneous Groups of Nucleosomes In Vivo. *Cell*. 160:1145–1158.

40. Shin, Y., Y.-C. Chang, D.S.W. Lee, J. Berry, D.W. Sanders, P. Ronceray, N.S. Wingreen, M. Haataja, and C.P. Brangwynne. 2018. Liquid Nuclear Condensates Mechanically Sense and Restructure the Genome. *Cell*. 175:1481-1491.e13.

41. Bray, A.J. 1994. Theory of phase-ordering kinetics. *Adv. Phys.* 43:357–459.

42. Cugliandolo, L.F. 2015. Coarsening phenomena. *Comptes Rendus Phys.*

43. Binder, K., and D. Stauffer. 1974. Theory for the Slowing Down of the Relaxation and Spinodal Decomposition of Binary Mixtures. *Phys. Rev. Lett.* 33:1006–1009.

44. Ahmad, S., F. Corberi, S.K. Das, E. Lippiello, S. Puri, and M. Zannetti. 2012. Aging and crossovers in phase-separating fluid mixtures. *Phys. Rev. E*. 86:061129.

45. Corberi, F., L.F. Cugliandolo, and H. Yoshino. 2011. Growing length scales in aging systems. In: Dynamical Heterogeneities in Glasses, Colloids, and Granular Media. *Oxford University Press*. pp. 370–406.

46. Brackley, C.A., J.M. Brown, D. Waithe, C. Babbs, J. Davies, J.R. Hughes, V.J. Buckle, and D. Marenduzzo. 2016. Predicting the three-dimensional folding of cis-regulatory regions in mammalian genomes using bioinformatic data and polymer models. *Genome Biol.* 17:59.

47. Lucas, J.S., Y. Zhang, O.K. Dudko, and C. Murre. 2014. 3D Trajectories Adopted by Coding and Regulatory DNA Elements: First-Passage Times for Genomic Interactions. *Cell*. 158:339–352.

48. Ponti, G., F. Palombi, D. Abate, F. Ambrosino, G. Aprea, T. Bastianelli, F. Beone, R. Bertini, G. Bracco, M. Caporicci, B. Calosso, M. Chinnici, A. Colavincenzo, A. Cucurullo, P. Dangelo, M. De Rosa, P. De Michele, A. Funel, G. Furini, D. Giammattei, S. Giusepponi, R. Guadagni, G. Guarnieri, A. Italiano, S. Magagnino, A. Mariano, G. Mencuccini, C. Mercuri, S. Migliori, P. Ornelli, S. Pecoraro, A. Perozziello, S. Pierattini, S. Podda, F. Poggi, A. Quintiliani, A. Rocchi, C. Scio, F. Simoni, and A. Vita. 2014. The role of medium size facilities in the HPC ecosystem: the case of the new CRESCO4 cluster integrated in the ENEAGRID infrastructure. In: 2014 International Conference on High Performance Computing & Simulation (HPCS). IEEE. pp.

1030–1033.

49. Allen, M.P., and D.J. Tildesley. 1989. Computer Simulation of Liquids. *Oxford University Press*.

50. Kremer, K., and G.S. Grest. 1990. Dynamics of entangled linear polymer melts: A molecular-dynamics simulation. *J. Chem. Phys.* 92:5057–5086.

51. Barbieri, M., S.Q. Xie, E. Torlai Triglia, A.M. Chiariello, S. Bianco, I. De Santiago, M.R. Branco, D. Rueda, M. Nicodemi, and A. Pombo. 2017. Active and poised promoter states drive folding of the extended HoxB locus in mouse embryonic stem cells. *Nat. Struct. Mol. Biol.* 24:515–524.

52. Annunziatella, C., A.M. Chiariello, S. Bianco, and M. Nicodemi. 2016. Polymer models of the hierarchical folding of the Hox-B chromosomal locus. *Phys. Rev. E*. 94:042402.

53. Annunziatella, C., A.M. Chiariello, A. Esposito, S. Bianco, L. Fiorillo, and M. Nicodemi. 2018. Molecular Dynamics simulations of the Strings and Binders Switch model of chromatin. *Methods*. 142:81–88.

54. Plimpton, S. 1995. Fast parallel algorithms for short-range molecular dynamics. *J. Comput. Phys.* 117:1–19.

55. Rosa, A., and R. Everaers. 2008. Structure and Dynamics of Interphase Chromosomes. *PLoS Comput. Biol.* 4:e1000153.

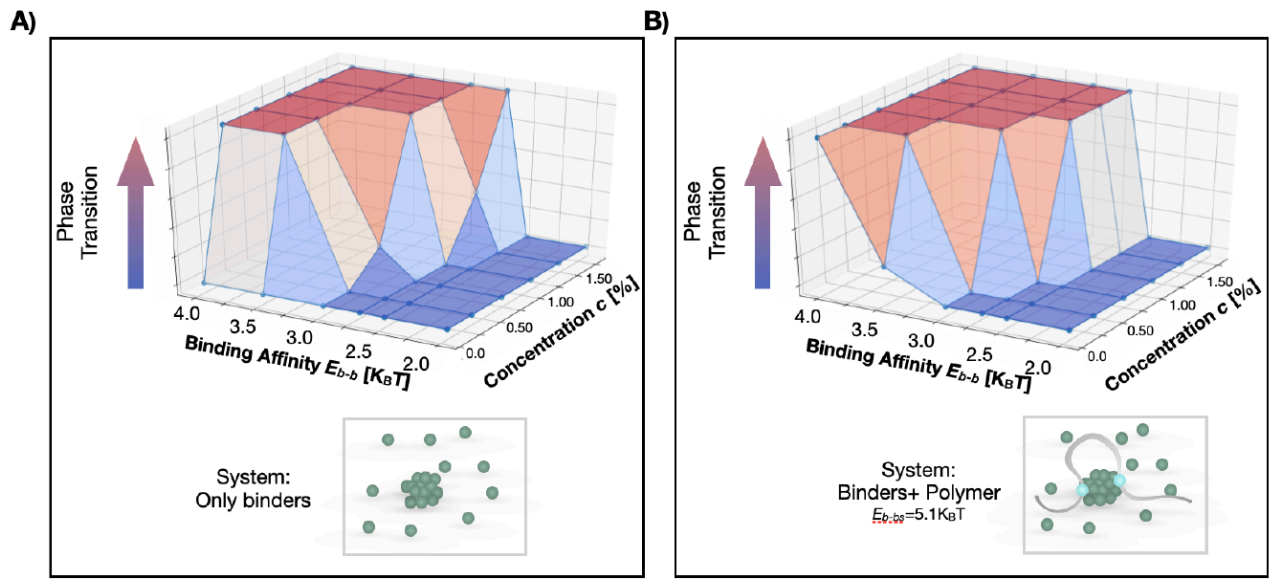

**Supplementary Figure 1: Phase diagram in different conditions**

Phase diagram can be influenced by the presence of the polymer. **A)** Phase diagram for the system composed only by the binders. Note that the diagram is very similar to the diagram of Figure 1E, where the system is composed by binders and a weakly interacting polymer. **B)** Phase diagram of the system composed by binders and a strongly interacting polymer, with a high affinity $E_{b-bs}$ = 5.1$K_B$T. Note that such interaction enhances the formation of phase-separated clusters for more combinations of control parameters.

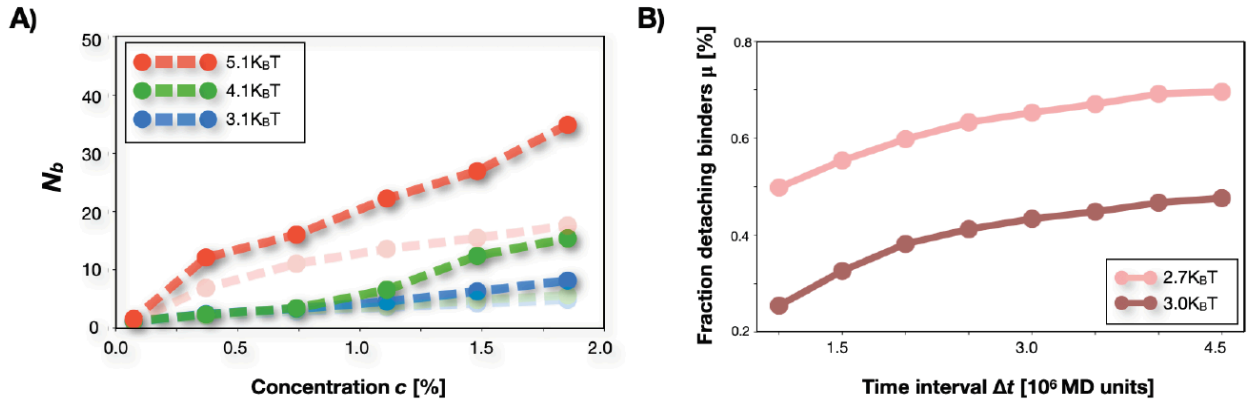

**Supplementary Figure 2: Influence of the binding sites and structural properties of the phase-separated clusters**

**A)** Number of binders $N_b$ in the largest cluster as a function of the concentration $c$, for different values of the affinity $E_{b\text{-}bs}$. The affinity $E_{b\text{-}b}$ between the binders is in the very weak range (2.5$K_B$T and 2.0$K_B$T for the opaque and transparent curves respectively), so to not induce phase-separation. Note that the presence of the binding sites induces a local increase in the concentration. **B)** Fraction of binders that detach from the cluster at least one time in a given time interval. For low affinities, the cluster exhibits highly dynamical properties since after enough time, which is comparable with the characteristic time between two contact events of regulatory elements, roughly 70% of the binders experience a detaching event.

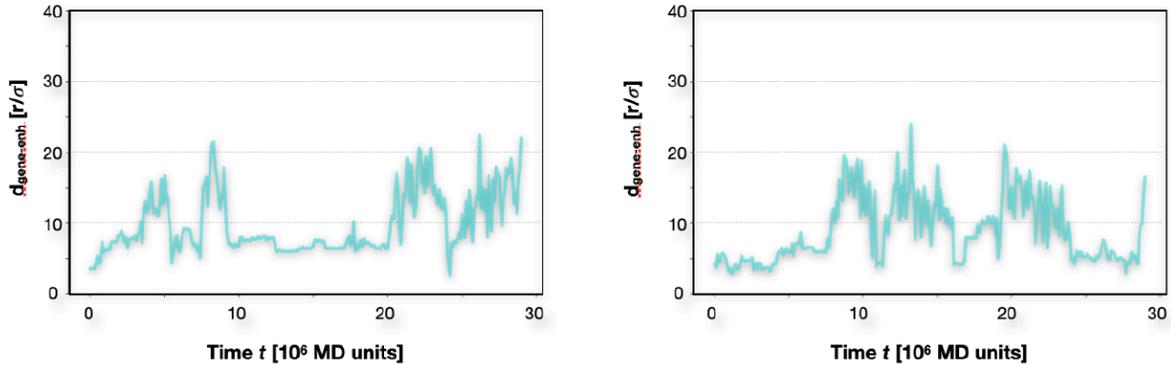

**Supplementary Figure 3: Gene-enhancer dynamics regimes are stable across changes of the system parameters**

Two independent examples of simulated gene-enhancer dynamics in the regime colored in cyan in Figure 3A. In this case, the binder diameter is $0.5\sigma$. Binding regions are made of 1 and 3 binding sites respectively, separated by an average linear distance of 50 beads. Affinity parameters ($E_{b\text{-}b}$ and $E_{b\text{-}bs}$) are the same used in Figure 3C, with a similar concentration per volume unit ($c \sim 1\%$). Note the stability of the contact with sharp increases of the gene-enhancer distance, due to detaching events of the polymer binding sites from the phase-separated cluster. In this respect, the behavior is analogous to the distance dynamics reported in Figure 3C.